# Laboratory Report (LR) to the paper

## Foundation of an analytical proton beamlet model for inclusion in a general proton dose calculation system

W. Ulmer[1,2] and B. Schaffner[1,3]

[1]Varian Medical Systems, Baden, Switzerland, [2]MPI of Biophysical Chemistry, Göttingen, Germany, [3]ETH Zürich

E-Mail: waldemar.ulmer@gmx.net

### 1. Integration of BBE and relationship to equations 1 - 4

Physical details and parameters with regard to BBE (Eq. (52)) are given elsewhere (ICRU 49). The critical behavior of BBE is the nonrelativistic limit, i.e. v = 0 (or E = 0), since it produces severe singularities, and a cutoff at E = 1 MeV is usually assumed in Monte Carlo codes. Therefore we only consider here the nonrelativistic case:

$$\left\{\begin{array}{l} -dE/dz = K \cdot \mathrm{v}^{-2} \cdot [\ln(2m \cdot \mathrm{v}^2 / E_I) + a_{shell} + a_{Barkas} + a_0 \mathrm{v}^2] \\ K = (Z\rho / A_N) \cdot 8\pi \cdot q^2 \cdot e^4 / 2m \end{array}\right\} \quad \mathbf{(L1)}$$

In order to avoid confusion, we note that *m* refers to the electron rest mass and *M* to the proton rest mass. Eq. (L1) can be integrated, if the following substitutions are carried out:

$$\left\{\mathrm{v}^2 = 2 \cdot E/M;\ \beta = 4 \cdot m/(E_{I.} \cdot M);\ E = \beta^{-1} \cdot \exp(-u/2)\right\} \quad (\mathrm{L2})$$

The application of Eq. (L2) to the logarithmic term of Eq. (L1) implies severe singularities, whereas the simultaneous integration of all terms of Eq. (L1) leads to mutual compensation of these critical terms and a cutoff is superfluous. A complete integration of Eq. (L1) with the help of Eq. (L2) and including relativistic correction terms (Bloch) yields the power expansions according to Eq. (1). With regard to water we obtain formula 2 (see also Tables L1 and L2). A comparison between ICRU49 data and formula 2 is shown in Fig. L1; the average standard deviation amounts to 0.13 %.

Table L1: Parameter values for Eq. (1) if $E_0$ is in MeV, $E_I$ in eV and $R_{CSDA}$ in cm

| $\alpha_1$ | $\alpha_2$ | $\alpha_3$ | $\alpha_4$ | $p_1$ | $p_2$ | $p_3$ | $p_4$ |
|---|---|---|---|---|---|---|---|
| $6.8469 \cdot 10^{-4}$ | $2.26769 \cdot 10^{-4}$ | $-2.4610 \cdot 10^{-7}$ | $1.4275 \cdot 10^{-10}$ | 0.4002 | 0.1594 | 0.2326 | 0.3264 |

Table L2: Parameter values for Eq. (2) if $E_0$ is in MeV, $E_I$ in eV and $R_{CSDA}$ in cm

| $a_1$ | $a_2$ | $a_3$ | $a_4$ |
|---|---|---|---|
| $6.94656 \cdot 10^{-3}$ | $8.13116 \cdot 10^{-4}$ | $-1.21068 \cdot 10^{-6}$ | $1.053 \cdot 10^{-9}$ |

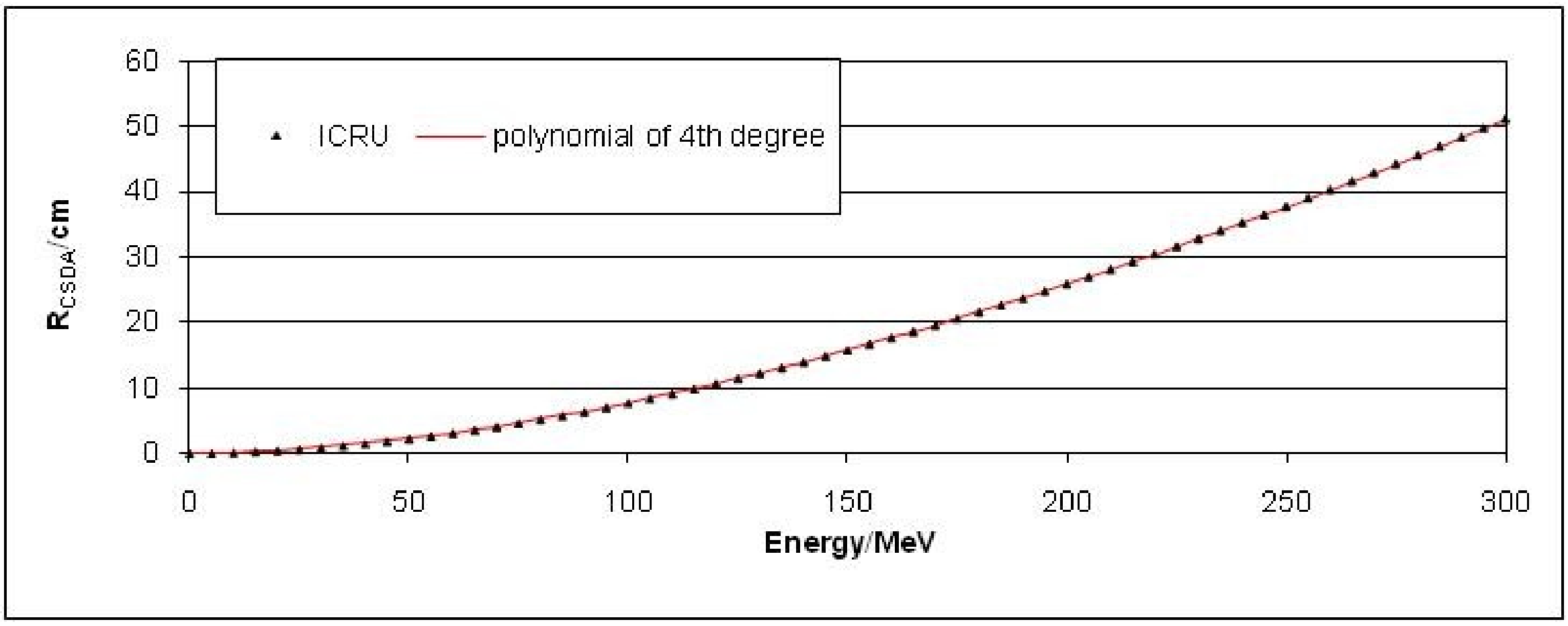

Fig. L1. Comparison of $R_{CSDA}$ between formula 2 and data in ICRU49.

A further possible way of integrating BBE leads to Gompertz-functions. A very low order approximation with the same accuracy as Eq. (2) is given by:

$$\boldsymbol{R_{CSDA} = 6.8469 \cdot 10^{-4} \cdot E_0 \cdot [1 + \sum_{k=1}^{N} (b_k - b_k \exp(-g_k \cdot E_0))]} \quad \textbf{(L3)}$$

The parameters $b_k$ (dimensionless) and $g_k$ of this equation with N = 2 are (Ulmer 2007); $b_1$ = 15.14450027, $b_2$ = 29.84400076, $g_1$ = 0.001260021 $MeV^{-1}$, $g_2$ = 0.003260031 $MeV^{-1}$. Eq. (L3) is also useful, since it can be inverted, i.e. $E_0 = E_0(R_{CSDA})$, whereas Eq. (2) leads to an ill-posed problem. The result is given by Eq. (3) and Table L3, if the restriction N = 5 to therapeutic protons is accounted for.

A through and back calculation of $R_{CSDA}(E_0)$ according to formula 2 in steps of 1 MeV up to 300 MeV and $E_0(R_{CSDA})$ according to Eq. (3) gave a mean standard deviation of 0.11 %. Eq. (L3) also provides to use the actual (residual) energy E(z) by the substitutions $R_{CSDA} \rightarrow R_{CSDA} - z$ and $E_0 \rightarrow E(z)$. Equations (3 - 8) are straightforward.

Table L3: Parameter values for the inversion Eq. (3) with N = 5, if $E_0$ is in MeV, $E_I$ in eV and $R_{CSDA}$ in cm (dimension of $A_k$: MeV/cm, $\beta_k$: cm). Note: Eq. (3) only requires $A_1,\ldots,A_5$ and $\beta_1,\ldots,\beta_5$.

| $A_1$ | $A_2$ | $A_3$ | $A_4$ | $A_5$ | $\beta_1$ | $\beta_2$ | $\beta_3$ | $\beta_4$ | $\beta_5$ |
|---|---|---|---|---|---|---|---|---|---|
| 99.639 | 25.055 | 8.8075 | 4.19001 | 9.1832 | 0.0975 | 1.24999 | 5.7001 | 10.6501 | 106.727 |
| $P_1$ | $P_2$ | $P_3$ | $P_4$ | $P_5$ | $q_1$ | $q_2$ | $q_3$ | $q_4$ | $q_5$ |
| -0.1619 | -0.0482 | -0.0778 | 0.0847 | -0.0221 | 0.4525 | 0.195 | 0.2125 | 0.06 | 0.0892 |

Eq. (3) can be extended from water to a medium with arbitrary density ρ, nuclear charge Z and mass number $A_N$ by the substitutions:

$$\left.\begin{aligned} A'_k &= A_k \cdot (18/10) \cdot Z \cdot \rho \cdot (75.1/E_I)^{q_k} \; /(A_N \cdot \rho_w) \\ \beta'_k &= \beta_k \cdot (10 \cdot \rho_w /18) \cdot (75.1/E_I)^{p_k} \cdot A_N \; /(\rho \cdot Z) \\ E(z) &= (R_{CSDA} - z) \cdot \sum_{k=1}^{5} A'_k \cdot \exp[-(R_{CSDA} - z)/\beta'_k\,] \end{aligned}\right\} (L4)$$

Note: $E_I$ of water amounts to 75.1 eV and $\rho_{water}$ to 1 g/cm$^3$; with respect to $A_{N,water}$ and $Z_{water}$ the Bragg rule has been applied, i.e., $A_{N,water}/Z_{water}$ = 18/10. Together with formula (L4), Eq. (3) incorporates the inversion of Eq. (1). With regard to heterogeneous media E(z) and dE(z)/dz can readily be determined in a stepwise manner, if $A_N$, ρ, and Z are known (details: Ulmer and Matsinos 2010).

The parameters of Eq. (7) and Table 1 solely refer to water. They have to be modified in similar fashion as by Eq. (L4):

$$\left.\begin{aligned} & c_p \Rightarrow c_p \cdot (18/10) \cdot Z \cdot \rho \cdot (75.1/E_I)^{q_k} \; /(A_N \cdot \rho_w) \;\; (k = 1,..,4) \\ & P_E^{\;-1} \Rightarrow P_E^{\;-1} \cdot (10 \cdot \rho_w /18) \cdot (75.1/E_I)^{\varepsilon} \cdot A_N \; /(\rho \cdot Z) \\ & \varepsilon = 0.0922 \end{aligned}\right\} \text{(L5)}$$

The powers $q_k$ (k = 1,..4) are identical with those of Table L3.

**2. Integration of the nuclear cross-section and decrease of the primary proton fluence**

In the LR of the publication (Ulmer 2007) we have given some information about a calculation model of the total nuclear cross-section proton – oxygen resulting from extended nuclear shell theory; a detailed elaboration can be found in a review (Ulmer and Matsinos 2010). The model is in very good agreement with measurement data of Chadwick et al 1996; the measurement data presented by Seltzer et al show some deviations between 30 MeV and 140 MeV.

Fig. L2 shows that for E < 7 MeV only elastic scattering processes at the oxygen nucleus occur, and E = 7 MeV is the threshold energy $E_{Th}$ due to the Coulomb barrier. The total nuclear cross-section has a resonance maximum at $E_{res}$ = 20.12 MeV and thereafter it decreases exponentially. In the domain E > 150 MeV and E < 270 MeV the asymptotic behavior is reached (the production of π-mesons requires proton energies E >> 270 MeV). Fig. L2 is also very closely related to the decrease of the primary proton fluence.

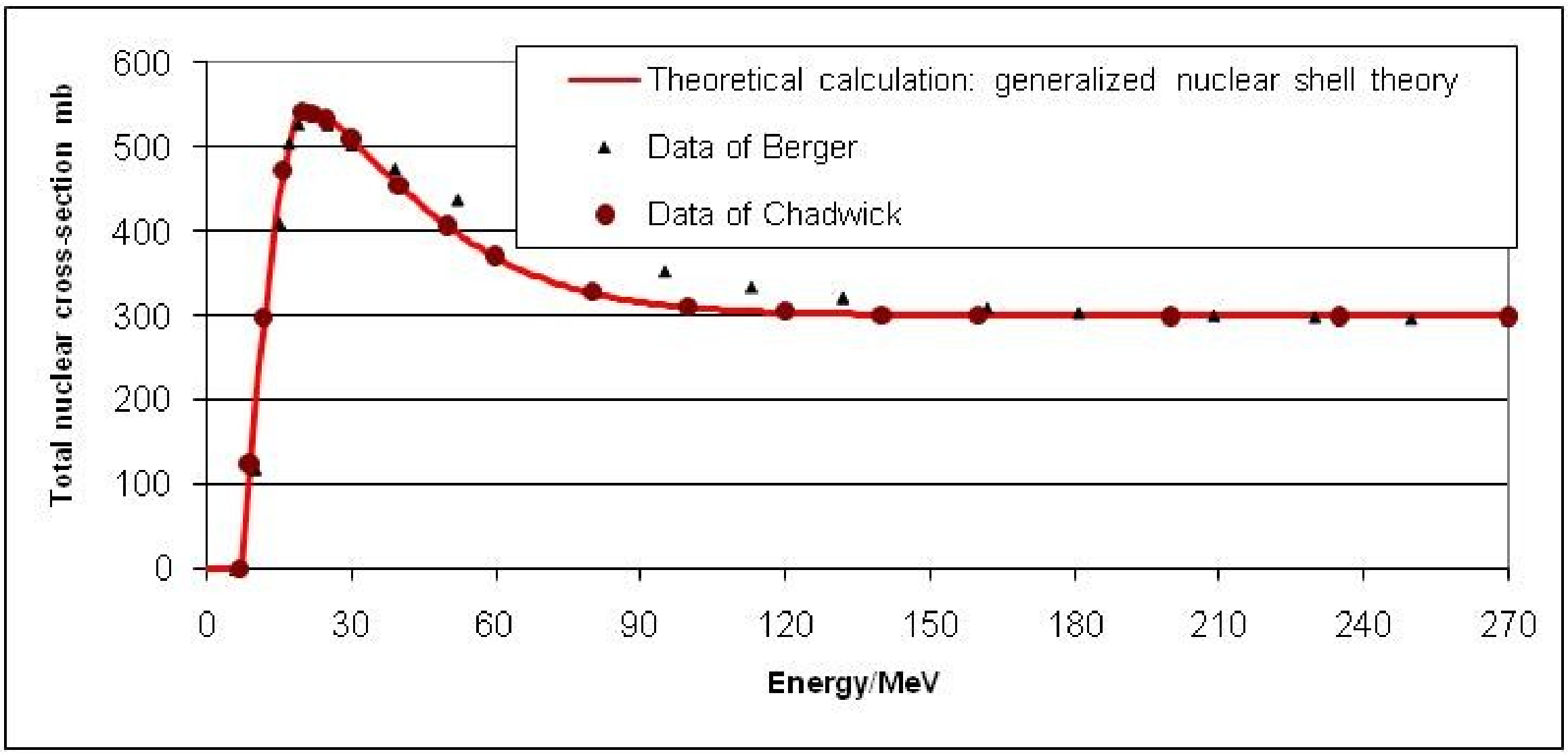

Fig. L2. Total nuclear cross-section of the proton - nucleus (oxygen) interaction.

By numerical evaluation, Boon (1998) presented the finding that in dependence of the primary proton energy the proton fluence decreases linearly. One would expect a jump, when a nuclear reaction is impossible, but due to statistical fluctuations this ‘jump’ is rather rounded, when the proton has lost most of its initial energy. An analytical integration (Ulmer 2007) also gave a linear correspondence, and the slope depends on $E_0$. There have been considered many attempts to fit solely this fluence decrease in dependence of $R_{csda}$ (see e.g. Bortfeld 1997 and references therein). Since the dependence of the difference $E_0$ - $E_{Th}$ has not been accounted for in these fits, they have been made artificially complicated and are not accurate. Eq. (4) is the result of an analytical integration with $\Phi_0$ as an arbitrary initial fluence. Fig. L3 summarizes the results of the previous analysis (Ulmer 2007); $\Phi_0$ is normalized to 1. For E < MeV the proton fluence again is constant without linear decrease. However, the range of 7 MeV protons is less than 1 mm and therefore only the roundness due to fluctuations can be verified.

In further communications (Ulmer and Schaffner 2006, Ulmer and Matsinos 2010) we have presented total nuclear cross-sections for materials different from the oxygen nucleus. They have some importance with regard to collimator scatter and passage of protons through bones. The fluence decrease of primary protons according to Figures L4 – L5 is analogous to that of water (Fig. L3). A comparison of Figure L5 with Figure L3 shows that the slope of the fluence decrease is steeper, in particular for proton energies < 100 MeV. For copper the threshold energy $E_{Th}$ now amounts to 8.24 MeV and for calcium 7.72 MeV, respectively. Due to the rather significant fluctuations we cannot verify a constant fluence at the end of the track. Eq. (8) can be applied to Fig. L5 with $E_{Th}$ = 8.24 MeV (copper) and $E_{Th}$ = 7.72 MeV (calcium). The power f for the computation of uq in Eq. (5) has to be modified (f = 0.755 (copper) and f = 0.86 (calcium)):

$$uq = \left(\frac{E_0 - E_{Th}}{M \cdot c^2}\right)^f \quad (L6)$$

$$f(Z, A_N) = a \cdot A_N^{-1} + b \cdot A_N^{-2/3} + c \cdot A_N^{-1/3} + d \cdot Z \cdot A_N^{-1/2} \quad (L7)$$

Formula (L7) is closely related to the nuclear collective model (Ulmer and Matsinos 2010), where the parameters *a, b, c* and *d* are interpreted in terms of different contributions to the total cross section. These parameters have been determined by this model: a = -0.087660001, b = -6.379250217 , c = 5.401490050and d = - 0.054279999.

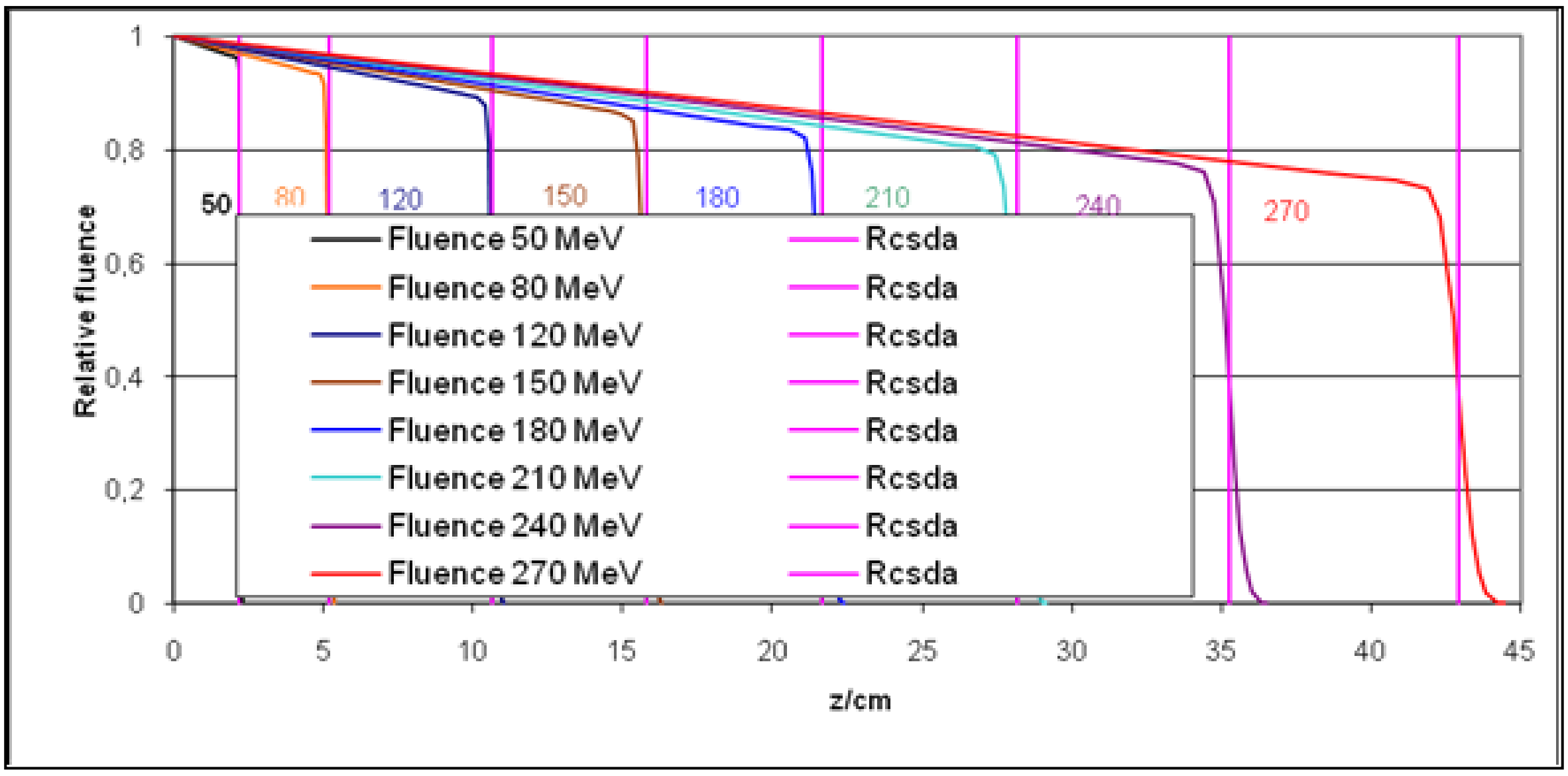

Fig. L3. Decrease of the primary proton fluence resulting from Fig. L2.

Formulas (8, 19) can be applied with regard to the modifications obtained by equations (1, L6 – L7). There are two applications, where Fig. (L5) is relevant: 1. Passage of protons through collimators. 2. Passage of protons through bone/metallic implants. In case 2, only a small path length has to be corrected. However, Figure L5 shows that the fluence decrease has also to be corrected according to the boundary conditions. Figures L3 and L5 do not yet provide information about the contributions $S_{sp,n}$ and $S_{sp,r}$. The case of nonreaction protons (nuclear potential/resonance scatter) has already been treated. According to Fig. 1 the contribution of reaction protons is particular important for E > 150 MeV with increasing energy.

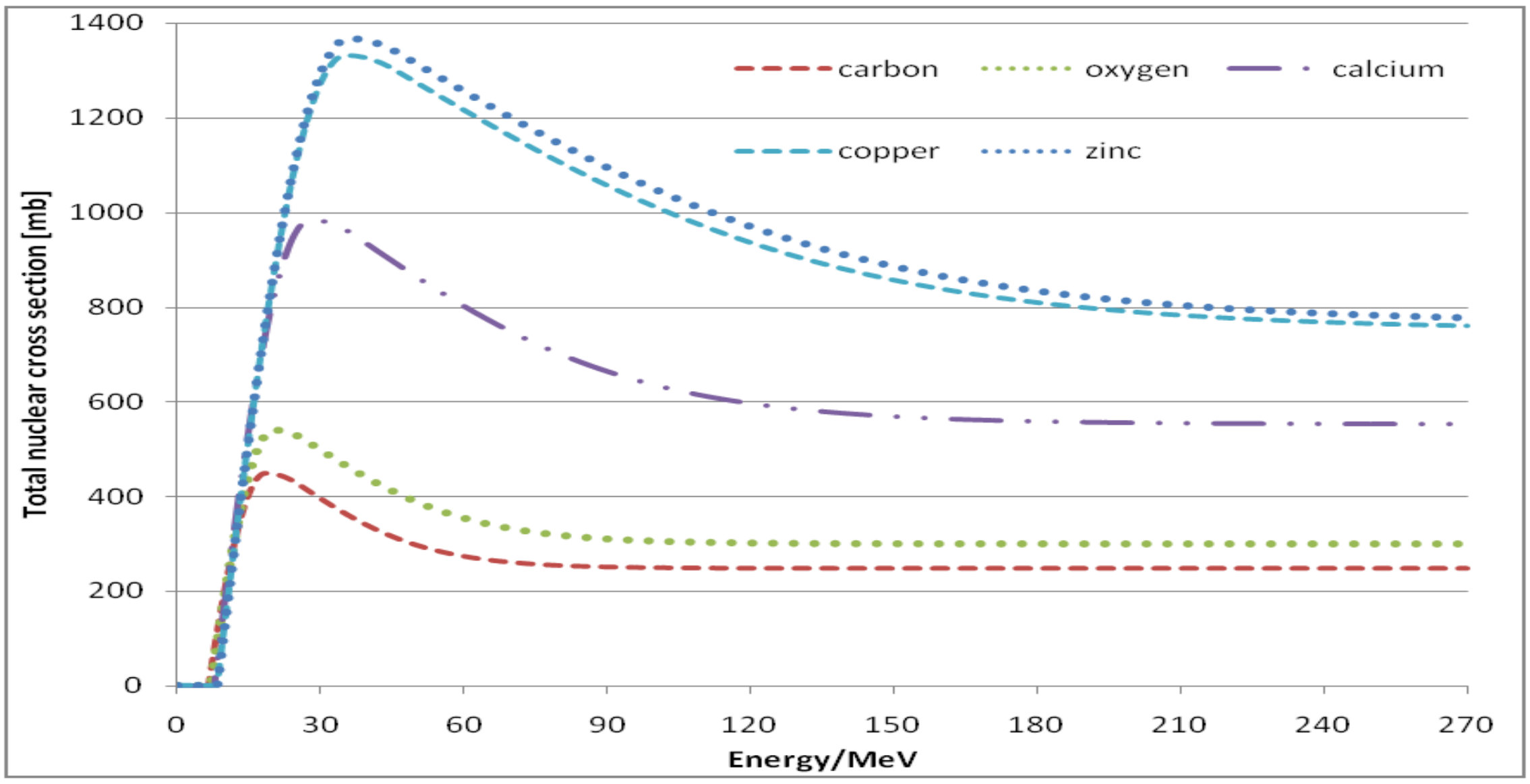

Fig. L4. Total nuclear cross-section of the proton interaction with the nuclei of some materials of interest (data from the Scientific Los Alamos library).

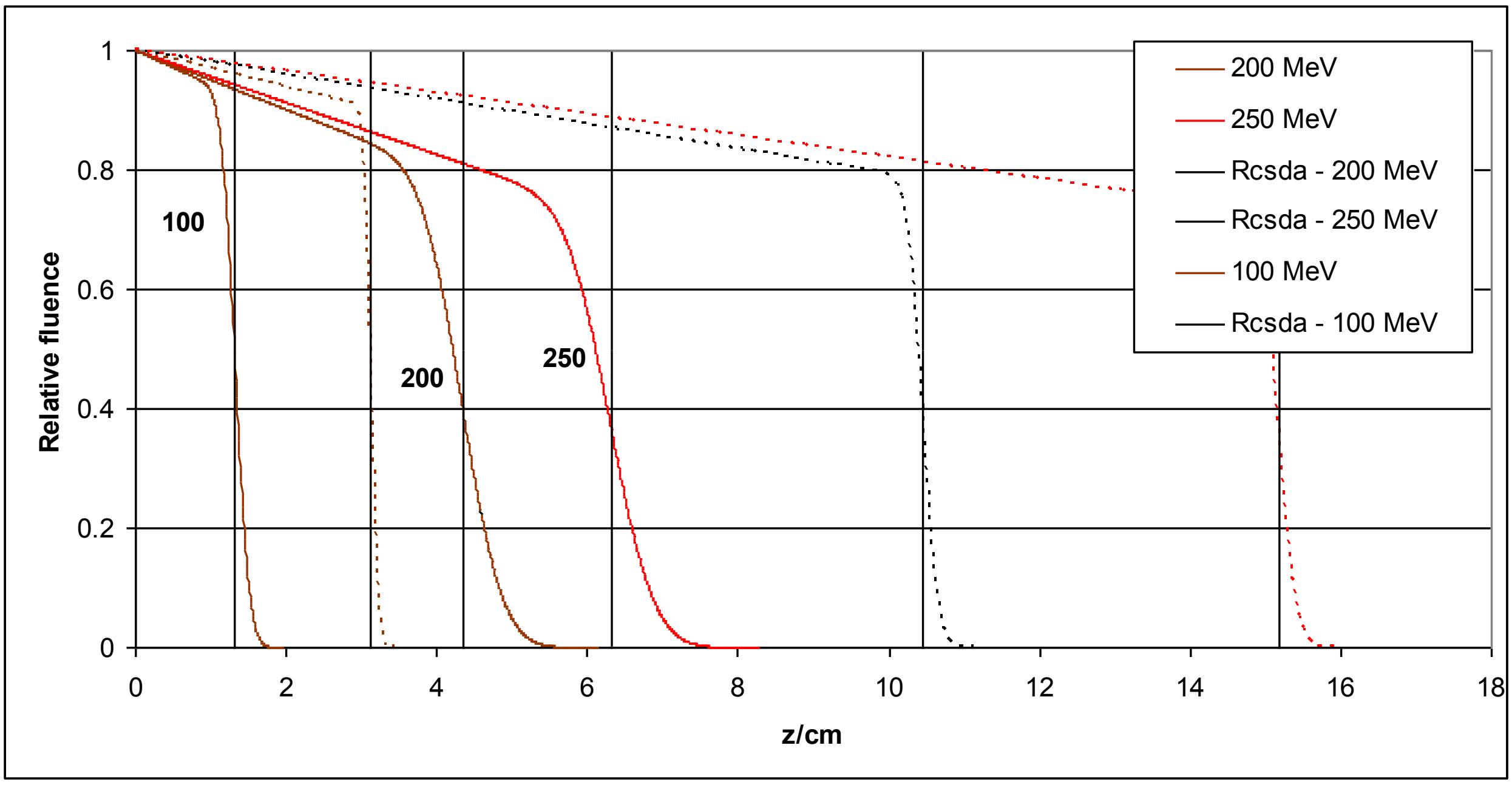

Fig. L5. Decrease of the fluence of 100, 200 and 250 MeV primary protons in brass (solid lines) and in Calcium (dashes). The $R_{CSDA}$ ranges are stated by perpendicular lines.

We now present the calculation formula for this case (detailed information is given in the review of Ulmer and Matsinos 2010). Thus $S_{sp,r}$ is proportional to $\Phi_0 \cdot 2 \cdot \upsilon \cdot C_{heavy}$ and a function *F*, depending on some further parameters. It should be mentioned that the parameters of Eq. (L8) are not restricted to the oxygen nucleus. From Figure L4 and Eq. (1) corresponding parameters of some further nuclei can be verified, e.g., $R_{csda}$, $E_{Th}$ and $E_{res}$. We use the following definitions:

$$\left.\begin{aligned}
&z_R = R_{CSDA}/\pi \\
&\tau_s = 0.55411;\ \tau_f = \tau_s - 0.000585437\cdot(E - E_{res}) \\
&\arg 1 = z/\sqrt{\tau_s^{\,2} + \tau_{in}^{\,2} + (R_{CSDA}/4\pi)^2} \\
&\arg 2 = (R_{CSDA} - z - \sqrt{2}\cdot\pi\cdot z_{shift})/\sqrt{\tau_f^{\,2} + (\tau_{straggle}/7.07)^2 + (R_{CSDA}/\sqrt{3}\cdot 4\pi)^2} \\
&\arg = (R_{CSDA} - 0.5\cdot z_{shift}\cdot\sqrt{\pi} - z)/(\sqrt{\pi}\cdot z_{shift})
\end{aligned}\right\}\quad (L8)$$

The result is the following connection:

$$\left.\begin{aligned}
&S_{sp,r} = (E_0/N_{abs})\cdot\Phi_0\cdot[2\cdot\upsilon\cdot C_{heavy}\cdot F_r + G]\cdot \\
&\cdot Q^{tot}{}_{as}(medium)\cdot A_N^{\,1/3}/(Q^{tot}{}_{as}(oxygen)\cdot A_{oxygen}^{\,1/3}) \\
&F_r = \varphi\cdot[\varphi_1 + \varphi_2\cdot\theta] \\
&\varphi = 0.5\cdot(1 + erf(\arg));\quad \varphi_1 = erf(\arg 1);\quad \varphi_2 = erf(\arg 2) \\
&\theta = \begin{pmatrix} e^{-1} - \exp(-(z - z_R)^2/z_R^{\,2})/(e^{-1} - 1) & (if\ z \le z_R) \\ 1 & (else) \end{pmatrix} \\
&G = (c_1\cdot z_{shift}\cdot\sqrt{\pi}/R_{CSDA})\cdot\exp(-(\tfrac{1}{\sqrt{\pi}}R_{CSDA} - z)^2/z_R^{\,2})
\end{aligned}\right\}\quad (L9)$$

Eq. (L9) requires the determination of the parameters given in Eq. (L8). All parameters are stated in the paper, excepted $E_{res}$. Thus for oxygen $E_{res}$ amounts to 20.12 MeV, but $E_{res}$ of some other materials (C, Ca, Cu, Zn) can be taken from Figure L4 or from a related calculation formula. With regard to the asymptotic cross-sections $Q^{tot}{}_{as}$ in Eq. (L9) we may either use Figure L4 or formulas, which may be taken in a detailed discussion of equations (L8 – L9) being published (Ulmer and Matsinos 2009, Ulmer and Matsinos 2010).

**3. The importance of Landau-Vavilov distributions in buildup of protons and the role of secondary (reaction) protons**

It is a noteworthy result (Ulmer 2007) that a Boltzmann distribution function (canonical ensemble) in connection with a Schrödinger equation (i.e. a nonrelativistic approach) for the energy transfer from protons to environmental electrons yields a Gaussian convolution. This connection can briefly be shown as follows:

Let Φ be a source function and φ an image function resulting from an exchange of particles and energy $E_{ex}$ and obeying a statistical distribution, then an exchange Hamiltonian H mutually couples the source field Φ (e.g. proton fluence) and the image field φ by the operator $F_H$:

$$\left.\begin{aligned} &\varphi = F_H \cdot \Phi = \exp(-H / E_{ex}) \cdot \Phi \\ &H = -\tfrac{\hbar^2}{2m} d^2 / dz^2 \\ &s^2 = 2\hbar^2 / m \cdot E_{ex} \end{aligned}\right\} \quad \text{(L10)}$$

$F_H$ is an operator representation of a canonical ensemble, and the Gaussian convolution kernel K is obtained as the Green's function by using the spectral theorem of operators:

$$\left.\begin{aligned} &F_H \cdot \exp(-i \cdot k \cdot z) / \sqrt{2\pi} = \exp(-k^2 \cdot s^2 / 4) / \sqrt{2\pi} \;\Rightarrow \\ &K = \tfrac{1}{2\pi} \cdot \int_{-\infty}^{\infty} \exp(-k^2 \cdot s^2 / 4) \exp(i \cdot k \cdot u) \exp(-i \cdot k \cdot z) dk = \\ &K = \tfrac{1}{s \cdot \sqrt{\pi}} \exp[-((u - z)/s)^2] \end{aligned}\right\} \quad \text{(L11)}$$

With the help K according to Eq. (L11) the connection between source field Φ and image φ can be represented in the more familiar fashion:

$$\varphi(s, z) = \int K(s,\, u - z) \cdot \Phi(u)\, du \qquad \text{(L12)}$$

The energy exchange of a proton with the environmental electrons corresponds to the transfer energy $E_{transfer}$. Only in the thermodynamical equilibrium, which results from further collisions of electrons, we can replace $E_{tranfer}$ by $k_B \cdot T$ ($k_B$: Boltzmann constant). This aspect is also the starting point of the calorimetric dosimetry. Equations (L10 – L11) directly lead to the density matrix formalism (von Neumann) and to an operator calculus of the quantum-statistical extension of the Feynman propagator theory. Instead of the spectral distribution resulting from the restriction to a simple plane wave representation exp(-i·k·z) it is also possible to make use of a complete power (Fourier) expansion (Ulmer 2007), i.e.:

$$\left\{\begin{aligned} &\tfrac{1}{\sqrt{2\pi}} \cdot \exp(-i \cdot k \cdot z) \Rightarrow \tfrac{1}{\sqrt{2\pi}} \cdot f(k) \exp(-i \cdot k \cdot z) \\ &f(k) = \sum_{n'=0}^{\infty} \alpha_{n'} \cdot k^{n'} \end{aligned}\right\} \quad \text{(L13)}$$

This extension provides more complex spectral distributions, if $F_H$ acts on the system of functions provided by Eq. (L13) and the integration over k according to Eq. (L11) has been carried out. It should also be noted that Eq. (L11) provides general forms of the density matrix and Feynman propagators and may be applied to the calculation of the stopping power (see also section 2.1). The transition to statistical mechanics via Green's function calculus yields an access to a Boltzmann equation and a connection to a paper of Sandison and Chvetsov (2000).

If the energy transfer $E_{transfer}$ is repeated n-times along the proton track (i.e. difference of the actual (or residual) proton energy E(n) – E(n-1)), then we have to substitute $s^2 \rightarrow n \cdot s^2$. This implies that $E_{transfer}$ is constant along the track. This is, however, not true in the environment of the Bragg peak, and s is itself z-dependent, i.e. s = s(z) (see also Eq. (7), where s(z) is replaced by $\tau_{straggle}$). It is evident that this way to account for statistical fluctuations of the local energy transfer $E_{transfer}$ is only valid in a nonrelativistic approach. Thus for protons with energy >> 50 MeV relativistic corrections are required, and only in the Bragg peak domain the significance of these corrections breaks down.

If we pass from the Schrödinger equation and Boltzmann distribution function to the relativistic Dirac Hamiltonian and the distribution function of Fermi-Dirac statistics, i.e., **α**, β: Dirac matrices, **p**: relativistic momentum:

$$\left.\begin{array}{l} \mathrm{H_D} = \alpha \cdot \mathrm{p} + \beta \cdot \mathrm{mc}^2 \\ \varphi = F_{H_D} \cdot \Phi = \frac{1}{\exp((\mathrm{H}_D - \mathrm{E}_F)/\mathrm{E}_{ex}} \cdot \Phi \end{array}\right\} \quad \text{(L14)}$$

$E_F$: energy of the Fermi edge, for the description of the fluctuations, we obtain according to Ulmer (2007) with the help of rather complicated calculations the convolution kernel $K_F$ and energy distribution function $S_E$. They have the shape:

$$K_F = N_f \cdot \sum_{l=0}^{\infty} B_l(n, mc^2) \cdot H_l((u - z - z_{shift}(l))/\sigma_n) \cdot \exp(-(u - z - z_{shift}(l))^2 / 2\sigma_n{}^2) \quad \text{(L15)}$$

$$\left\{\begin{array}{l} S_E = N_f \cdot \exp(-(E_n(k) - E_{Average,n})^2 / 2\sigma_E(n)^n)) \cdot \\ \cdot \sum_{l=0}^{\infty} b_l(n, mc^2) \cdot (E_n(k)/2E_{transfer,n})^l \\ E^2 = m^2c^4 + \hbar^2 k^2 c^2 \end{array}\right\} \quad \text{(L16)}$$

$N_f$ is a normalization factor and the repetition factor *n* has the same meaning as above. Note that $E_{Average,n}$ results from the energy of the Fermi edge $E_F$, and by restricting Eq. (L16) to the lowest order, i.e., $l = 0$ and all higher order terms are omitted, Bohr's classical formula of energy fluctuations is obtained.

Now $E_{Average}$ and $z_{shift}$ depend in every order on the energy of the Fermi edge $E_F$. The expansion coefficients $b_l$ and $B_l$ are determined by a uniformly convergent expansion of the sech-function in terms of a Gaussian and Hermite polynomials $H_l$. However, for the purpose of therapeutic protons we do not need fully exploit these expansions. In a low order approximation, equations (L15– L16) imply a Vavilov distribution, containing Landau tails. For very low energies (nonrelativistic limit: $E \rightarrow 0$), the limit of a Gaussian kernel is obtained again. In particular, a low-level approximation of Eq. (L16) is of interest for analytical convolutions (‘generalized Gaussian kernel’) to account for deviations of symmetrical fluctuations. A glance to Fig. L6 shows that $E_{max}$ of the energy transfer of 250 MeV protons amounts to ca. 617 keV and decreases along the proton track. The straight-line represents the nonrelativistic determination of $E_{max}$. The contributions $I_{Lan1}$ and $I_{Lan2}$ in section 2.6 represent integrations of terms up to the order 2 of Eq. (L15).

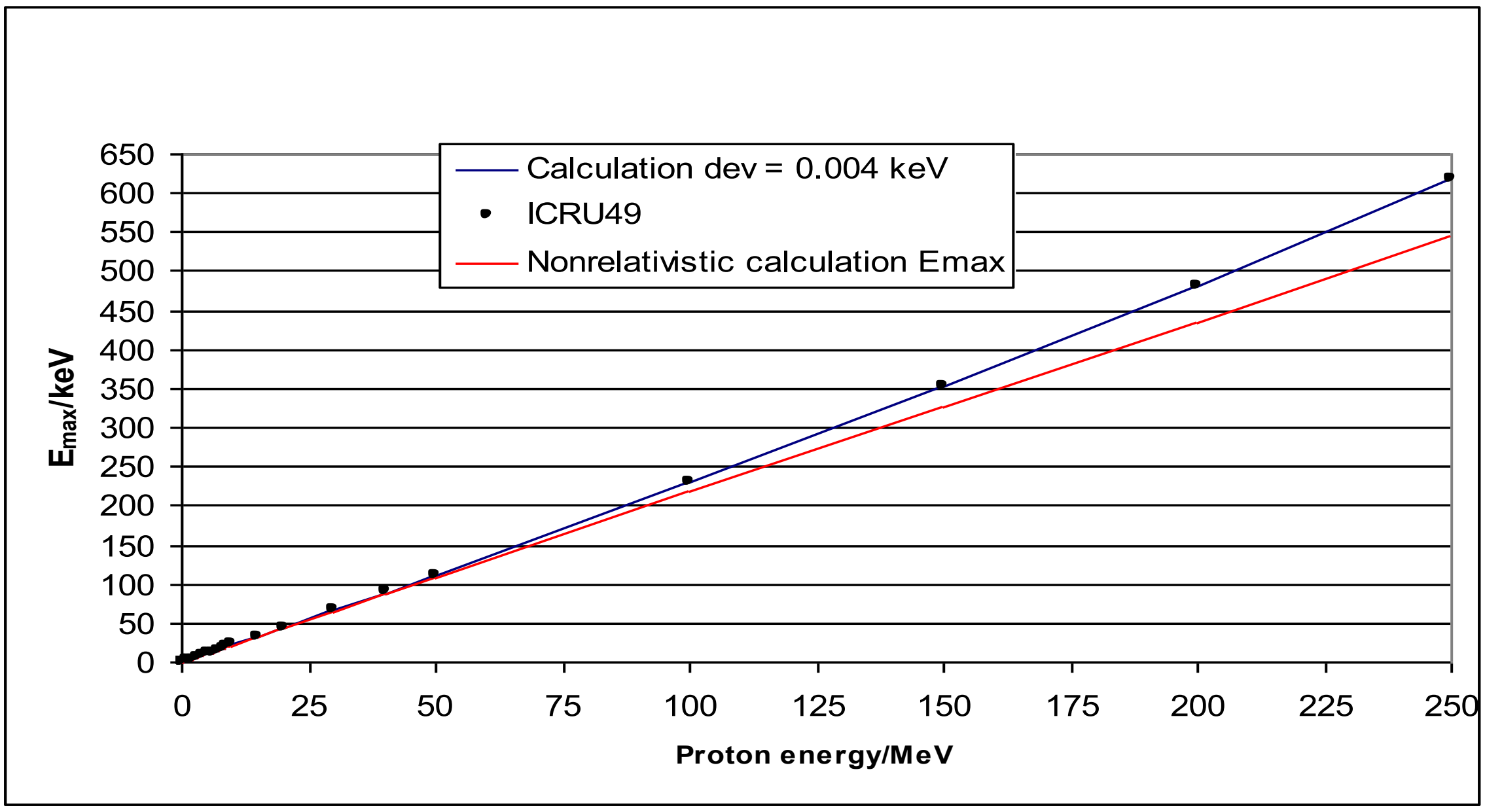

Fig. L6. Maximum transfer energy $E_{max}$ of protons in dependence of the energy (dots: ICRU49). The straight line below indicates a linear connection between proton energy and maximal energy transfer resulting from a nonrelativistic calculation.

In order to obtain a fast approximation up to order 2 of Landau tails according to equations (28 – 29), we have assumed that in the environment of the impinging surface the first and second order terms are leading, whereas in the domain of decreased proton energies only terms up to order 1 are significant (they are certainly also of interest at surface). All necessary calculation parameters have been determined by fits to the more general theory as outlined above and by fits to Monte Carlo (GEANT4), since the code has also the option to use a model of Vavilov distributions.

In the following, we abbreviate $\tau_{Lan1}$ with $\tau_1$ and $\tau_{Lan2}$ with $\tau_2$; $N_{f1}$ and $N_{f2}$ are suitable proportionality factors . The determination of formula 28 results from the first order Hermite polynomial. In principle, we have to evaluate the convolution integral

$$I_{Lan1} = N_{f_1} \cdot \frac{2}{\sqrt{\pi}} \cdot \frac{1}{\tau_1 \cdot R_{csda}} \cdot \int_0^{R_{csda}} (u - z_{Ls}) \cdot \exp[-(u - z - z_{Ls})^2 / \tau_1^2] \quad \text{(L17)}$$

This integral can be evaluated by the substitution:

$$u' = (u - z - z_{Ls}) / \tau_1 \qquad \text{(L18)}$$

$$I_{Lan\,1} = N_{f_1} \cdot \frac{2}{\sqrt{\pi} \cdot R_{csda}} \cdot \int_{-(z_{Ls}+z)/\tau_1}^{(R_{csda}-z-z_{Ls})/\tau_1} (u' \cdot \tau_1 + z) \cdot \exp(-u'^2) du' \quad \text{(L19)}$$

Eq. (L19) simply yields error functions and Gaussians, which we write in the form $I_{Lan1} = T_{erf} + T_{gauss}$:

$$T_{erf} = N_{f_1} \cdot \frac{z}{R_{csda}} \cdot [erf(z + z_{Ls}) / \tau_1) + erf((R_{csda} - z - z_{Ls}) / \tau_1)] \quad \text{(L20)}$$

$$T_{gauss} = N_{f_1} \cdot \tau_1 / (\sqrt{\pi} \cdot R_{csda})[\exp(-(z + z_{L_s})^2 / \tau_1^2)) - \exp(-(R_{csda} - z - z_{L_s})^2 / \tau_1^2))] \quad \text{(L21)}$$

The Hermite polynomial of second degree (unnormalized) reads:

$$H_2(\xi) = 4 \cdot \xi^2 - 2 \qquad \text{(L22)}$$

We have simply to evaluate

$$I_{Lan2} = 4 \cdot N_{f_2} \cdot \frac{2}{\sqrt{\pi}} \cdot \frac{1}{\tau_2} \cdot R_{Lan2}^{-2} \cdot \int_0^{R_{Lan2}} (u - z_{Ls})^2 \cdot \exp(-(u - z - z_{Ls})^2 / \tau_2^2) \, du + T_c \quad \text{(L23)}$$

The term $T_c$ refers to the convolution of the factor '- 2' in $H_2$ according to relation L22, which is only a 'standard problem'. We replace $\tau_1$ by $\tau_2$ in the substitution L18: this yields the following evaluation

$$\left.\begin{aligned} I_{Lan2} &= 4 \cdot N_{f_2} \cdot \tfrac{2}{\sqrt{\pi}} \cdot R_{Lan2}{}^{-2} \cdot \int_{-(z+z_{Ls})/\tau_2}^{(R_{Lan2}-z-z_{Ls})/\tau_2} (u'^{\,2} \cdot \tau_2{}^2 + z^2 + 2 \cdot u' \cdot z) \cdot \exp(-u'^{\,2})\, du' + T_c \\ T_c &= -2 \cdot N_{f_2} \cdot \tfrac{2}{\sqrt{\pi}} \cdot \int_{-(z+z_{Ls})/\tau_2}^{(R_{Lan2}-z-z_{Ls})/\tau_2} \exp(-u'^{\,2})\, du' \end{aligned}\right\} \quad \text{(L24)}$$

Thus the terms containing $z^2$, $2 \cdot u' \cdot z$ and $T_c$ have already been solved. The remaining term containing $u'^2 \cdot \tau_2^2$ can be integrated with the help of the chain rule, i.e.:

$$\int_a^b u'^2 \cdot \exp(-u'^2)\, du' = -0.5 \cdot (b \cdot \exp(-b^2) - a \cdot \exp(-a^2)) + T_{erf}(a,b) \quad \text{(L25)}$$

The integration boundaries a, b are stated in Eq. (L26).

$$\left.\begin{aligned} T_{erf} &= 0.5 \cdot \tfrac{\sqrt{\pi}}{2} \cdot [erf(b) - erf(a)] \\ a &= (z + z_{Ls}) / \tau_2 \,;\; b = (R_{Lan2} - z - z_{Ls}) / \tau_2 \end{aligned}\right\} \text{(L26)}$$

It seems noteworthy to add some comments referring to equations (L17 – L26). In equations (28 – 29) we have stated not all terms as required by the above evaluation. The reason is that we have carried out a selection and accounted only for those terms, which lead to significant contributions. Terms of the order < 0.2 % have been neglected. This is justified, since the overall contributions of Landau tails are small in the energy domain of therapeutic protons and saving computation time should also be accounted for.

Carlsson et al (1977) have reported a buildup of the Bragg curve of 185 MeV protons ($E_{max}$ of the incident beam: ca. 446 keV) and interpreted this effect as a sole result of secondary protons, since these protons are rather generated along the proton track than at surface. Following aspects should be noted:

1. We have already verified that secondary reaction protons significantly contribute to the buildup of monoenergetic protons. With regard to polychromatic protons the spectral parameter $\tau_{in}$ implies an enhancement of the buildup by accounting for the Landau tail. Further modulations are possible by collimator scattering occurring in broad proton beams and depending on the field-size.

2. The question also arises, whether the transport of δ-electrons is responsible for the buildup similar

to the Compton scatter of photons. However, $E_{max}$ of 250 MeV protons amounts to 617 keV (Fig. L6) and with regard to the cited 185 MeV protons this energy is lower. The range of these electrons is too small to explain a buildup. In order to produce $E_{max}$ of the order 4 MeV, the proton energy should amount to ca. 1 GeV.

3. A possible contribution of the γ-quanta resulting from β+-decays of heavy recoils and γ-quanta by the annihilation of positrons cannot be significant, since the interaction processes are rare enough.

In agreement with GEANT4 the depth dose curve of primary protons (250 MeV) shows a valley in the middle part of the plateau resulting from the corresponding fluence decrease (Fig. 2). If the transport of all secondary protons is included, the total depth dose curve does not show this valley. A comparison with the results of Medin et al (1997) is noteworthy. If the transport of secondary protons is only partially accounted for or even omitted (PTRAN), this valley can also be recognized at the total depth dose curve. In addition, it is necessary to include the 'secondaries' in an accurate manner. According to the cited authors PTRAN leads for 200 MeV protons to a dose contribution of ca. 10 % at the depth z = 20 cm, whereas very early theoretical calculations of Zerby et al (1965) provided 17 % at the depth z = 20 cm. This is in agreement with Fig. L3 and may be calculated with the help of section 2.5. In order to be consistent with the total nuclear cross-section (Fig. L2), the resulting Fig. L3 and the classifications in section 2.5, we consider all protons as secondary protons, which result from the total nuclear cross-section. The so-called 'reaction protons (*sp,r*)', which stand in a close relationship to the heavy recoils, are very dominant for E > 150 MeV with increasing tendency. Their depth dose curve certainly shows a maximum along their track, but not a typical Bragg peak due to the broad spectral distribution (Fig. 1). The secondary (nonreaction) protons lead to a Bragg peak shifted slightly to a lower energy due to further losses of energy (resonance scatter). In various publications on therapeutic protons this distinction has not clearly been pointed out.

In relationship to figures 1 – 4, which deal with the role of reaction protons and the Landau tail of 250 MeV monoenergetic protons, we present some further calculations of primary protons (monoenergetic) and comparisons with GEANT4 calculations (Fig. L7). If we glance at this figure, which shows complete Bragg curves, we are hardly able to verify significant differences between GEANT4 and theoretical calculations. Therefore, we additionally present the buildup domain (see Fig. L8) and the behavior of the Bragg curves in the region with z > 15 cm (Fig. L9). Thus Fig. L8 clearly shows the small influence of the asymmetric energy transfer (Landau tail) to the buildup in the high energy domain. Figures L9 and L10 yield an indication of the role of cutoffs in GEANT4 due to the severe singularities in the environment of the distal end. As already pointed out (section LR.1 and Ulmer 2007) these singularities emerge from the behavior of BBE in the low energy domain, i.e., if either E → 0 or if the logarithmic term becomes 0. By that, the singularity problem is a general feature of Monte-Carlo codes. Resulting from the energy/range straggling ($\tau_{straggle}$) and the cutoff, the peak height calculated by GEANT4 is slightly lower than by the presented theoretical calculation. In all other remaining domains the difference between GEANT4 and theoretical calculation is extremely small, i.e., less than 0.2 %. With reference to Figure L8 it should be mentioned that the small buildup effect due to the Landau tail disappeared in GEANT4, if the Vavilov distributions was reduced to a Gaussian one as mentioned in section 2.6.

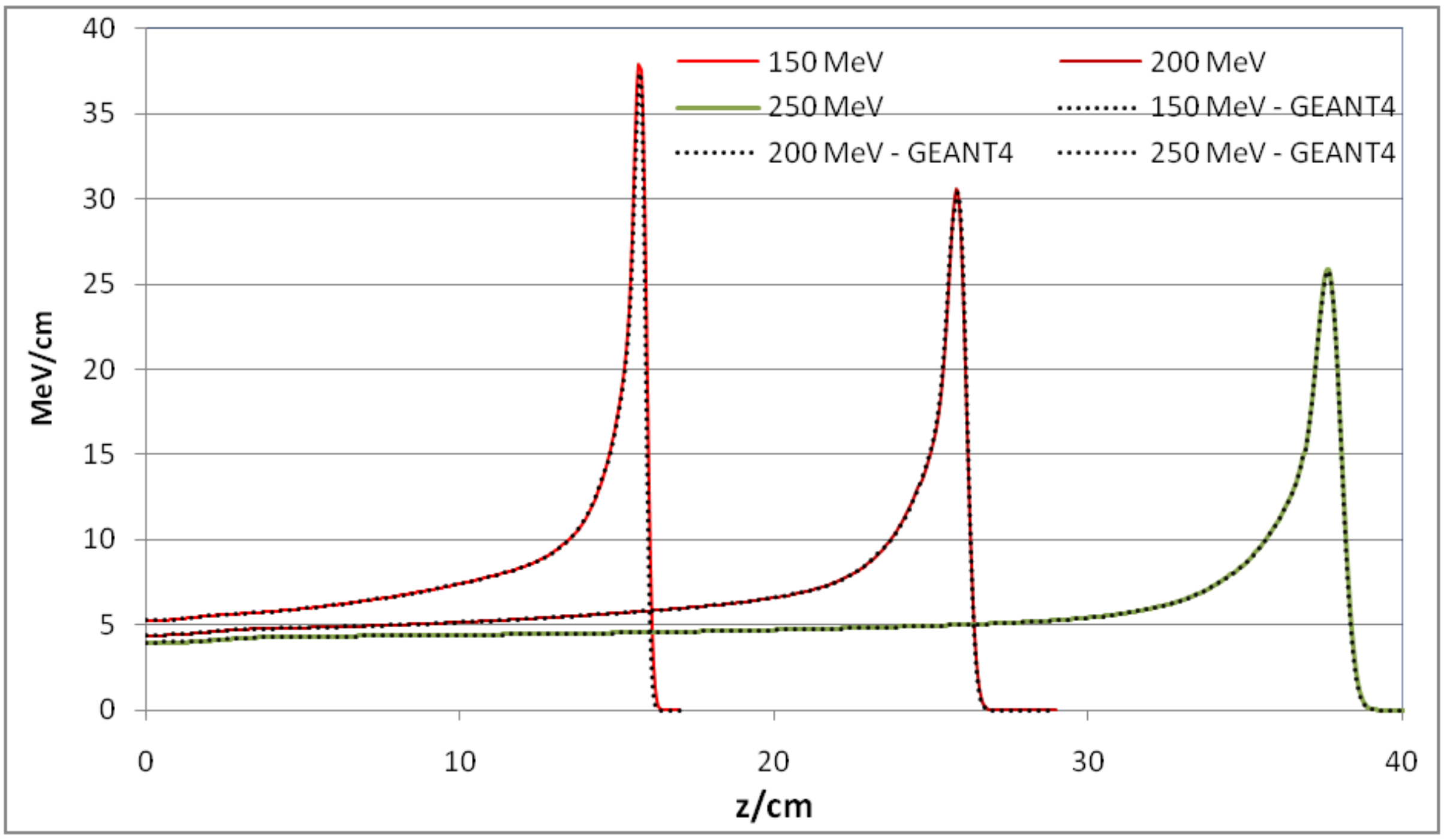

Fig. L7. Comparison of calculated stopping powers of monoenergetic primary protons with GEANT4 (dots).

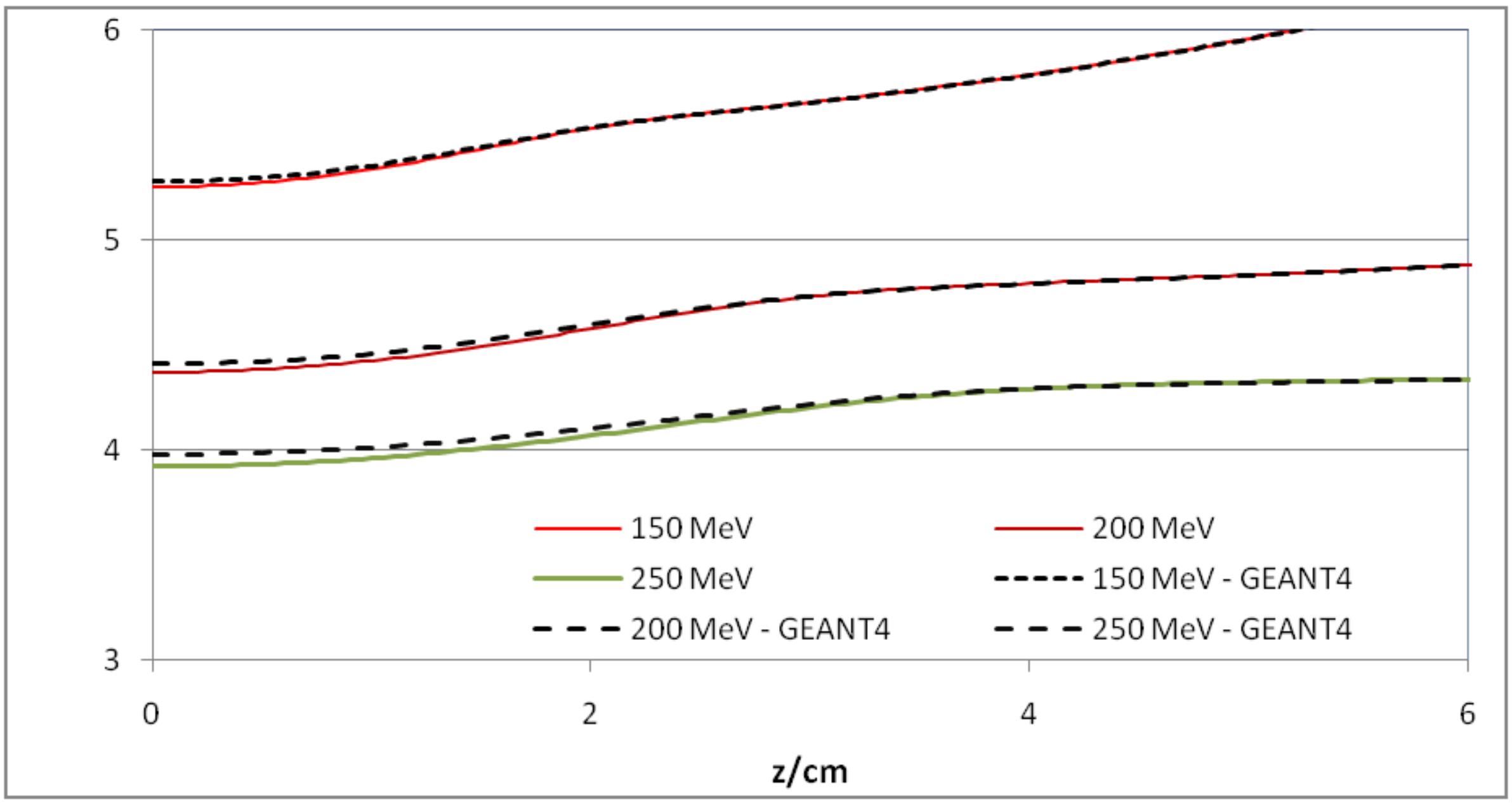

Fig. L8. Section of Fig. 7 in the buildup region.

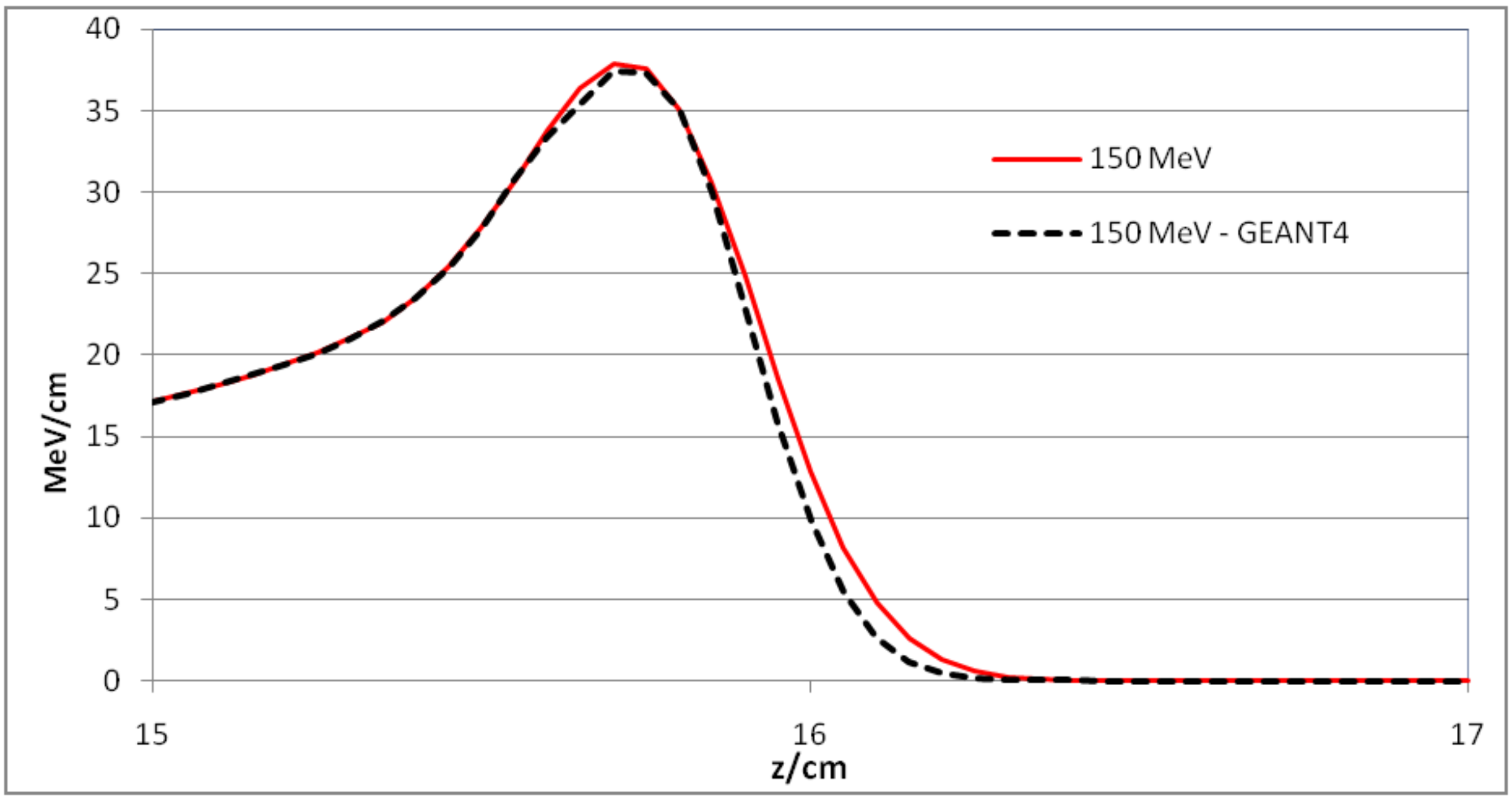

Fig. L9. Section of Fig. L7 in order to indicate the differences at the distal end.

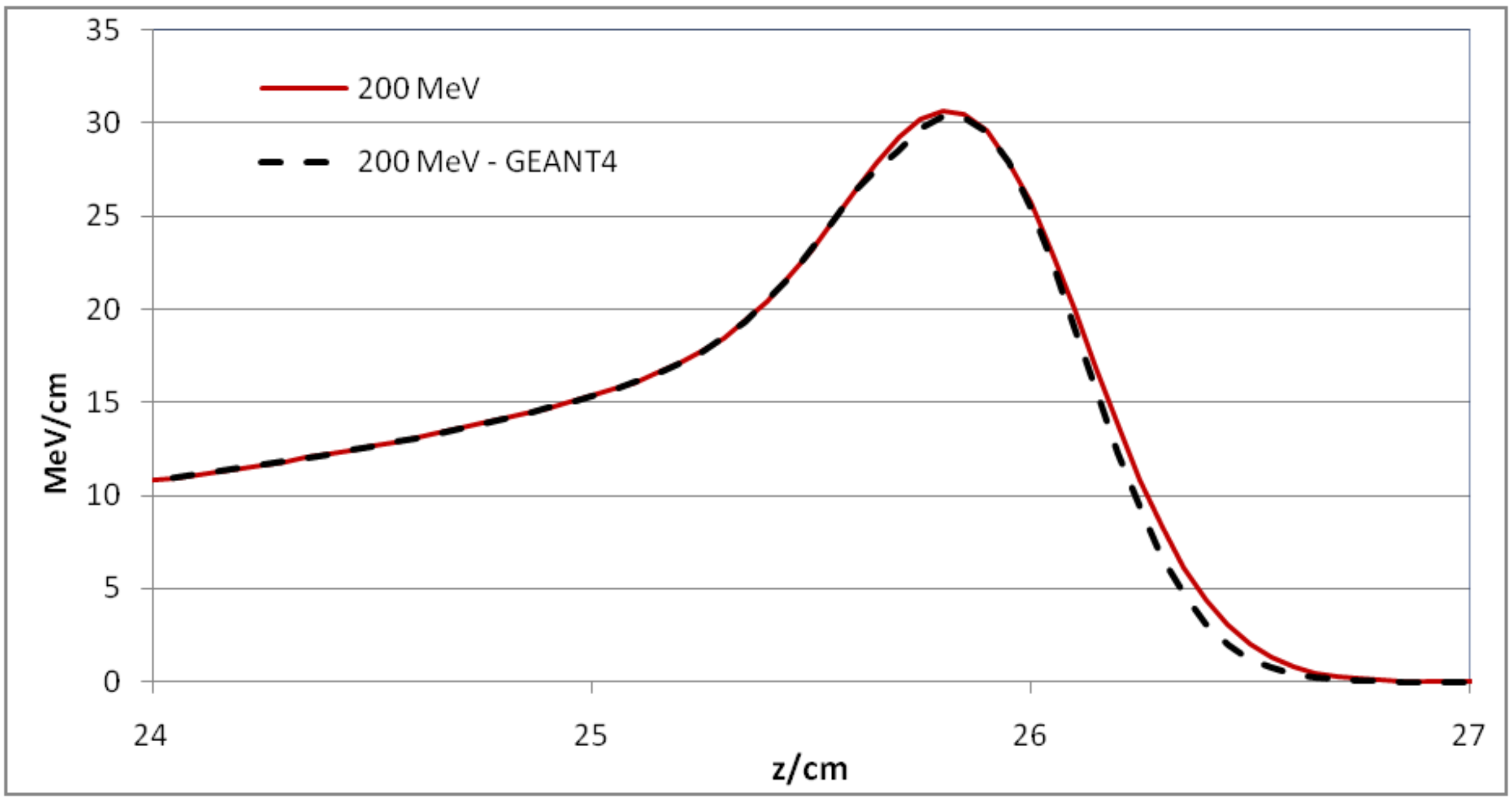
200 MeV
200 MeV - GEANT4
MeV/cm
z/cm
35
30
25
20
15
10
5
0
24
25
26
27

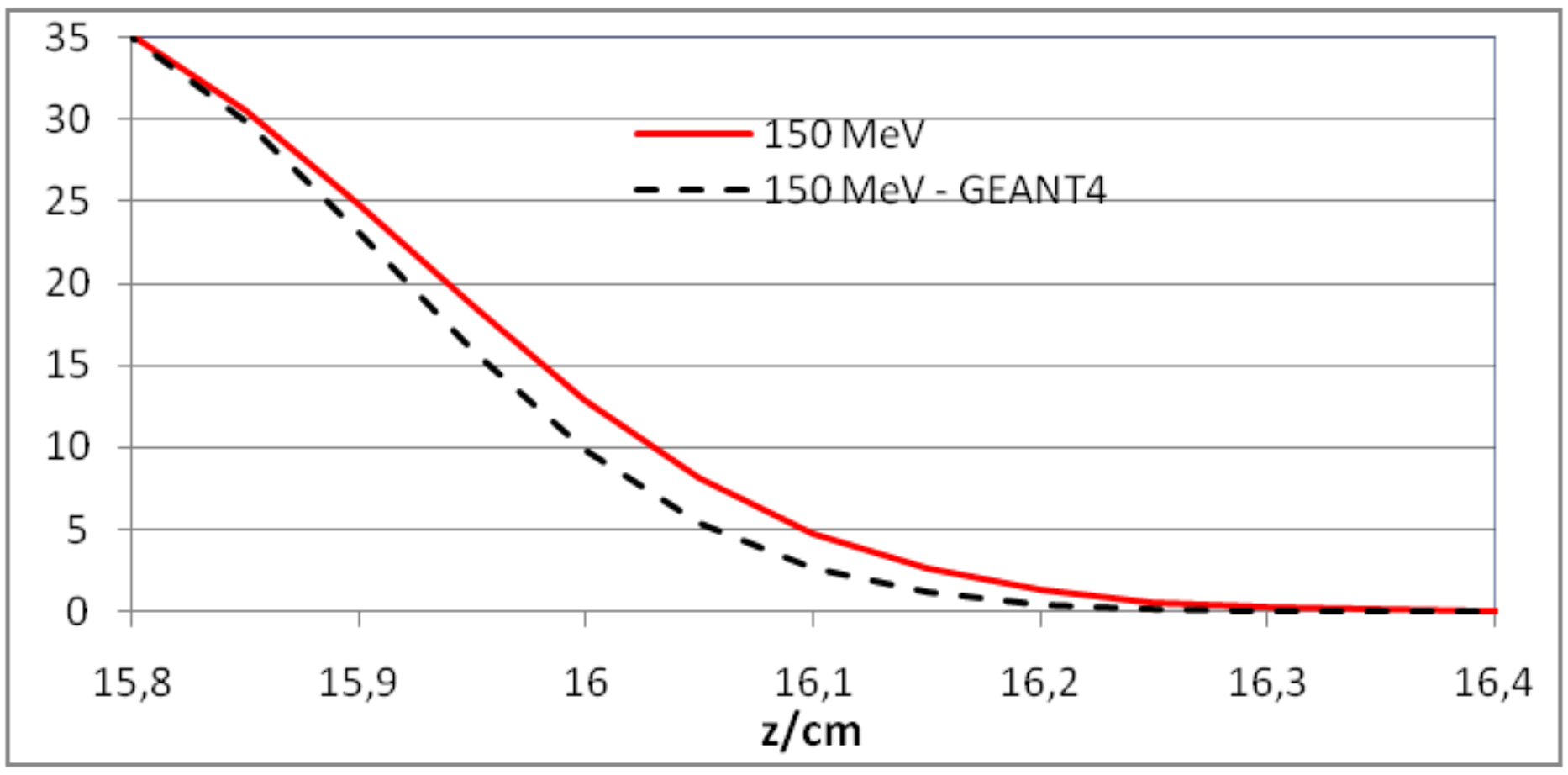
150 MeV
150 MeV - GEANT4
z/cm
35
30
25
20
15
10
5
0
15,8
15,9
16
16,1
16,2
16,3
16,4

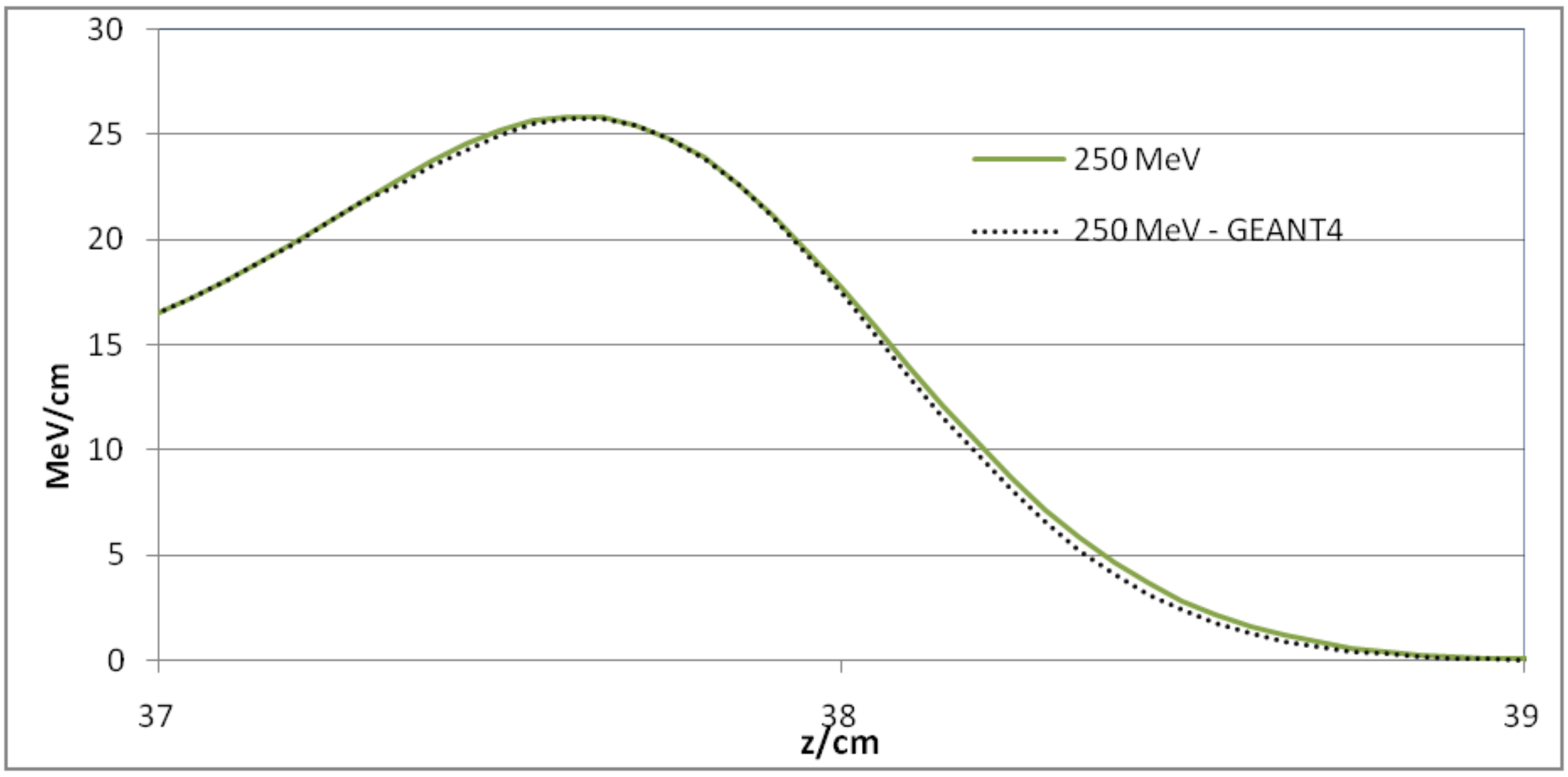
250 MeV
250 MeV - GEANT4
MeV/cm
z/cm
30
25
20
15
10
5
0
37
38
39

Fig. L10. 150 MeV protons at the distal end (solid: calculation, dashes: GEANT4).

**4. Multiple scatter procedure of primary and secondary protons**

In the present implementation of Eclipse, the lateral scatter of protons is treated by an approximate version of the multiple scatter theory (Bethe 1953, Molière 1955, Gottschalk et al 1993, Gottschalk et al 1999). However, we improved the Highland approximation (Highland 1995 and 1997) by the use of two Gaussian kernels in order to describe the lateral tail of primary protons and secondary (nonreaction) protons (*sp,n*) more accurate. GEANT4 indicated that two Gaussian kernels are sufficient for primary protons. Thus the first Gaussian accounts for the inner part of multiple Molière scatter, which is steeper than the Highland approximation, whereas the second Gaussian has a much larger half-width to approximately describe the tail. The Highland approximation assumes a slightly broader half-width than necessary for the inner part in order to account partially for the tail. For secondary (reaction) protons (*sp,r*), we restricted the lateral kernel to one modified Gaussian due to the significantly smaller contribution of them. Thus, the calculation model of the lateral kernel for water is given by:

$$K_{lat,prim}(r,z) = \frac{C_0}{\pi\tau_{lat}(z)^2}\cdot e^{-\frac{r^2}{\tau_{lat}(z)^2}} + \frac{(1-C_0)}{\pi\tau_{lat,LA}(z)^2}\cdot e^{-\frac{r^2}{\tau_{lat,LA}(z)^2}} \qquad \text{(L27)}$$

We use a weight coefficient $C_0$ of 0.96 for the contribution of the main Gaussian. The calculation of $\tau_{lat}(z)$ (inner part) and $\tau_{lat,LA}(z)$ (large angle) is carried out as follows:

$$\left.\begin{aligned}
&\tau_{lat}(z) = f\cdot 0.626\cdot\tau_{max}\cdot Q(z)\\
&f = 0.9236\\
&Q(z) = \frac{e^{\frac{1.61418\cdot z}{R_{CSDA}}}-1}{e^{1.61418}-1}\cdot 0.5\cdot\left[erf\left(\frac{R_{CSDA}-z}{\tau_{distal}}\right)+1\right]\\
&\tau_{max} = (E_0/176.576)^p;\ p = 1.5+\begin{cases}0.00150\cdot(176.576-E_0), & \text{if } E_0\le 176.576\\ 0.03104\cdot\sqrt{E_0-176.576}, & \text{if } E_0>176.576\end{cases}
\end{aligned}\right\} \qquad \text{(L28)}$$

The parameter $\tau_{distal}$ will be discussed at the end of this section. A good approximation for a model with a single Gaussian for primary protons can be obtained from the above equations by substituting $C_0$ and f by 1. In agreement with results of Gottschalk et al 1993 the change from water to other materials is obtained by scaling of Q(z) with the help of a different value for $R_{CSDA}$. The error function term models the Gaussian distribution of the stopping distribution due to range straggling. The protons which have undergone only very small angle scattering are closer to the central axis of the beamlet and travel slightly further.

A fit of Monte Carlo results shows that $\tau_{lat,LA}(z)$ can be determined by the kernel:

$$\tau_{lat,LA}(z) = \frac{0.90563}{1-e^{\frac{-1}{0.252}}} \cdot \tau_{\max} \cdot \left[ e^{-\frac{(1-z/R_{CSDA})^2}{0.252}} - e^{-\frac{1}{0.252}} \right] \cdot 0.5 \cdot \left[ erf\left( \frac{R_{CSDA}-z}{\tau_{distal}} \right) + 1 \right] \qquad \text{(L29)}$$

The scaling properties of Q(z) still hold, since only the ratio $z/R_{CSDA}$ enters Eq. (L29). For the secondary reaction protons and heavy recoils, we use separate Gaussian kernels, i.e.:

$$\left. \begin{aligned} \tau_{sp,r}(z) &= \tau_{max} \cdot Q(z) \\ \tau_{heavy}(z) &= \tau_{heavy}(E(z)) \end{aligned} \right\} \qquad \text{(L30)}$$

Q(z) and $\tau_{max}$ are the same as defined in Eq. (L28), but $\tau_{heavy}(E(z))$ is given by the same Eq. (24) as above except that it depends on the local energy E(z) instead of $\tau_{heavy}(E_0)$.

It might appear that Eq. (L27) and the calculation of $\tau_{max}$ according to Eq. (L28) are just obtained by fitting methods. However, the spatial behavior of the multiple scatter theory can be written by a Gaussian and Hermite polynomials $H_{2n}$:

$$\left. \begin{aligned} K(r,z,lat) &= N \cdot e^{-r^2/lat(z)^2} \cdot \begin{pmatrix} a_0 + a_2 \cdot H_2\left(\frac{r}{lat(z)}\right) + a_4 \cdot H_4\left(\frac{r}{lat(z)}\right) \\ + a_6 \cdot H_6\left(\frac{r}{lat(z)}\right) + 0\ (higher\ \ order\ \ terms) \end{pmatrix} \\ N &= 1/(\sqrt{\pi} \cdot lat(z)); \quad r^2 = x^2 + y^2 \end{aligned} \right\} \qquad \text{(L31)}$$

The parameters of the Hermite-polynomial expansion are: $a_0 = 0.932$; $a_2 = 0.041$; $a_4 = 0.019$; $a_6 = 0.008$. It is now the task to determine a linear combination of two Gaussians according to Eq. (L27) with different half-widths by a least square-fit (standard task in many problems). This way, we are able to go beyond the Highland approximation.

The calculation of $\tau_{max}$ is carried out in the following way: The differential cross-section is given by:

$$\left. \begin{aligned} \frac{dlat(z)^2}{dz} &= \frac{dlat_E^{\ 2}}{dE \cdot \frac{dE}{dz}} \\ \frac{dlat_E^{\ 2}}{dE} &= \frac{\alpha_{Mediaum}}{E^2} \end{aligned} \right\} \qquad \text{(L32)}$$

In our calculations, we only used $\alpha_{Water}$. $\alpha_{Medium}$ is proportional to $Z \cdot \rho / A_M$, i.e. nuclear charge, mass density and mass number. An evaluation of E(z) and dE/dz in Eq. (L32) above can be simplified, when using the Bragg-Kleeman rule:

$$\left.\begin{aligned} R_{CSDA} &= 0.00259 \cdot E_0{}^p \\ R_{CSDA} - z &= 0.00259 \cdot E(z)^p \end{aligned}\right\} \quad \text{(L33)}$$

The inversion of Eq. (L33) is given by:

$$E(z) = \left(\frac{R_{CSDA} - z}{0.00259}\right)^{1/p} \quad \text{(L34)}$$

An accurate application of this rule requires consideration of p depending on $E_0$. p is in the order of 1.7 – 1.8 according to Ulmer (2007). With the help of Eq. (L34) dE(z)/dz can be computed and the evaluation of Eq. (L32 ) yields:

$$lat_{\max} = \sqrt{\int_0^{R_{CSDA}} \left[\frac{\alpha_{Water}}{E(z)^2}\right] \cdot \frac{dE(z)}{dz} \cdot dz} \quad \text{(L35)}$$

One might expect that the lateral scatter functions according to equations (L28 – L29) continuously increase up to z = $R_{csda}$. However, this assumption would be valid, if the energy spectrum for scattered protons would be identical at a depth z independent of the scatter angle and fluctuations due to the energy/range straggling $\tau_{straggle}$ and $\tau_{in}$, respectively. In fact, there are small fluctuations of the lateral scatter functions along the proton tracks. In particular, from Bragg peak down to the distal end there is a significant difference between those protons, which have only undergone small angle scatter in this domain, and those with larger scatter angle. The latter protons have deposited their energy in an oblique path, and therefore they stop earlier and cannot reach the distal end of the Bragg curve. It is clear that the scatter functions for primary and secondary protons ($\tau_{lat}$, $\tau_{lat,LA}$ and $\tau_{sp}$) depend on $\tau_{straggle}$ and $\tau_{in}$, which are the origin of these fluctuations and yielding the significant change of the energy spectrum at the end of the proton tracks. In order to describe this behavior by a mathematical model, we prefer to use a Gaussian convolution, which is certainly justified in the domain of Bragg peak (low energy region of the proton tracks). As an example we use the function Q(z) in Eq. (L28), which determines both $\tau_{lat}$ and $\tau_{sp}$. We denote the fluctuation parameter by $\tau_{distal}$; the connection to straggle parameters will be considered thereafter.

The crude model assumes:

$$Q_0 = \begin{cases} \frac{\exp(1.161418 \cdot z / R_{CSDA}) - 1}{\exp(1.161418) - 1} \quad (if \; z \le R_{CSDA}) \\ 1 \quad (if \;\; z > R_{CSDA}) \end{cases} \qquad \text{(L36)}$$

The more realistic model taking all the arguments with regard to fluctuations into account is obtained by:

$$Q(z) = \frac{1}{\tau \cdot \sqrt{\pi}} \cdot \int_{-\infty}^{R_{csda}} Q_0(\xi) \exp[-(z-\xi)^2 / \tau_{distal}{}^2] d\xi \qquad \text{(L37)}$$

Using methods, often applied in this paper, we obtain the modified Eq. (L28) for Q(z) instead of $Q_0(z)$ according to Eq. (L37):

$$Q(z) = \frac{\exp(1.161418 \cdot \; z / R_{csda}) \; - 1}{\exp(1.161418) \; - 1} \cdot \frac{1}{2} \cdot [1 + erf((R_{csda} - z) / \tau_{distal})] \qquad \text{(L38)}$$

This result implies that Q(z) increases exponentially along the proton track, as long as the error-function is 1. Only in the environment of z = $R_{csda}$ Q(z) decreases rapidly. However, this depends actually on τ. The connection between τ and the above stated convolution parameters for energy/range straggling is a principal question. One might assume that for proton pencil beams with energy/range straggling we can set:

$$\tau_{distal}{}^2 = \tau_{straggle}{}^2 + \tau_{in}{}^2 \qquad \text{(L39)}$$

This assumption might be reasonable, since the proton history resulting from the beam-line and expressed by $\tau_{in}$ should be accounted for. As already pointed out we have made use of GEANT4 results for the adaptation of the scatter functions with regard to monoenergetic protons. It turned out that the best adaptation in the domain from the Bragg peak to the distal end can be obtained, if we put for monoenergetic protons:

$$\tau_{distal}{}^2 = \tau_{straggle}{}^2 + \tau_{Range}{}^2 \qquad \text{(L40)}$$

The mean standard deviation for protons with $E_0$ = 50 MeV up to 250 MeV in intervals of 25 MeV amounts to 2.6 %. $\tau_{Range}$ is given by Eq. (17). Therefore it may be justified to modify Eq. (L40) by:

$$\tau^2 = \tau_{\text{straggle}}{}^2 + \tau_{\text{in}}{}^2 + \tau_{\text{Range}}{}^2 . \qquad \text{(L41)}$$

$\tau_{Range}$ (Eq. (17)) is also a convenient parameter to include the initial phase space of beamlets, since the radial energy dependence of a beamlet (i.e. decrease of the energy) implies varying ranges relative to the central ray. With regard to the evaluation of all scanning data of an Accel machine the corrections by fitting $\tau_{Range}$ to account for initial phase space properties amounted to $\tau_{Range,fit} = \tau_{Range} (1 + \varepsilon)$ with $\varepsilon < 0.12$. Thus by the substitution $\tau_{Range} \rightarrow \tau_{Range,fit}$ a slight correction of Eq. (L41) is obtained.

An important aspect is the handling of the above formulas in the case of media different from water. According to findings of Gottschalk et al (1993) we are able to represent lateral scatter functions in a dimensionless manner approximately valid for arbitrary media by the substitutions $\tau' = \tau_{lat}/\tau_{max}$ and $s = z/R_{csda}$, i.e., $\tau' = \tau'(s)$ should be independent on the medium. Gottschalk et al demonstrated that this relationship approximately holds even for lead. Figure L11 presents the findings of Gottschalk et al and incorporates an essential tool with regard to scatter in heterogeneous media. An evaluation of $\tau_{max}$ with tools worked out in a review (Ulmer and Matsinos 2010) provided the scaling factors: $\tau_{max}$(calcium) = $\tau_{max}$(water)·1.0281; $\tau_{max}$(copper) = $\tau_{max}$(water)·1.0394; $\tau_{max}$(lead) = $\tau_{max}$(water)·1.0826.

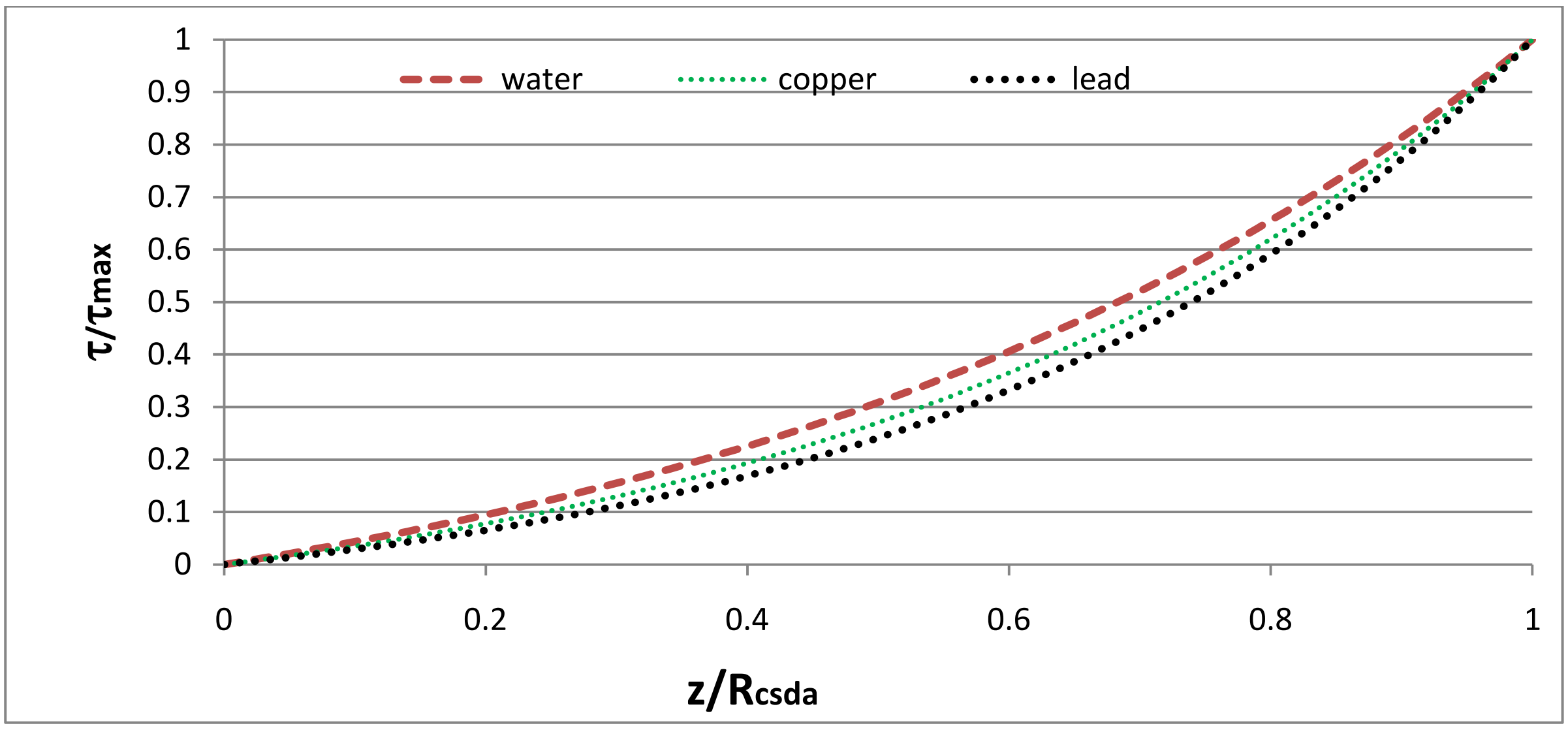

Fig. L11. Scaled representation of the lateral scattering function of water, copper and lead (Gottschalk et al 1993)

**5. Calculation of the absolute normalization factor**

The normalization factor $N_{abs}$, which appears in all contributions, provides normalization of the Bragg curves to the absolute dose in MeV/cm (in water). All distances are stated in units of cm and the energy in MeV. $N_{abs}$ is important with regard to the monitor unit calculation, *A* is the area integral of the complete integral over S(z), if S(z) is given by relative values (1 or 100 % at the Bragg peak). All necessary integrals and can be carried out analytically.

$$\left.\begin{aligned} &A = \int_{-\infty}^{\infty} S(z)\,dz \\ &N_{abs} = \frac{A}{E_0} = \frac{A_{pp1} + A_{pp2} + A_{sp,n} + A_{sp,r} + A_{rp}}{E_0} \end{aligned}\right\} \quad (L42)^{++}$$

$^{++}$In order to make apparent the integration procedure we state an example by Eq. (L48 – L49) at the end of this section. The recoil protons are treated here as primary protons. The integral over the contribution of primary protons is split into two parts for computational reasons. $A_{pp1}$ is calculated under the assumption, that there is no loss of primary protons, i.e. $\Phi_{pp} = \Phi_0$.

$$\left.\begin{aligned} A_{pp1} = \; & C_1\sqrt{\pi}\tau_{straggle} + 2\cdot C_2\cdot R_{csda} + 2\cdot C_3\cdot\frac{z_{max}}{PE\cdot\pi}\cdot\left[1-e^{-\frac{PE\cdot\pi\cdot R_{csda}}{z_{max}}}\right] \\ & +\frac{2}{3}\cdot C_4\cdot R_{csda} + C_{Lan1}\cdot E_0^{-1}\cdot R_{Lan1} - \frac{4}{3}\cdot C_{Lan2}\cdot R_{Lan2} \end{aligned}\right\} \quad (L43)$$

$A_{pp2}$ represents the energy lost by the primary protons due to nuclear reactions:

$$A_{pp2} = -\upsilon\cdot\frac{uq}{R_{csda}}\cdot\left\{\begin{aligned} &\frac{1}{2}\cdot C_1\cdot\tau_{straggle}\cdot\left(\sqrt{\pi}\cdot\left(1+erf\left(\frac{R_{csda}}{\tau_{straggle}}\right)\right)+\tau_{straggle}^{\;2}\cdot e^{-\frac{R_{csda}^{\;2}}{\tau_{straggle}^{\;2}}}\right) \\ &+C_2\cdot R_{csda}^{\;2} + 2\cdot C_3\cdot\left(\frac{z_{max}}{PE\cdot\pi}\right)^2\cdot\left[\frac{PE\cdot\pi\cdot R_{csda}}{z_{max}} - 1 + e^{-\frac{PE\cdot\pi\cdot R_{csda}}{z_{max}}}\right] \\ &+\frac{1}{2}\cdot C_4\cdot R_{csda}^{\;2} + E_0^{-1}\cdot C_{Lan1}\cdot R_{Lan1}^{\;2} - \frac{1}{2}\cdot C_{Lan2}\cdot R_{Lan2}^{\;2} \end{aligned}\right\} \quad (L44)$$

$A_{sp}$ is obtained by an integration of the contribution of secondary protons $S_{sp}$. $A_{sp,n}$ is somewhat smaller than $A_{pp2}$, since it does not include the energy carried away by photons or neutrons or used in nuclear reactions.

$$A_{sp,n} = \upsilon\cdot\frac{uq}{R_{csda}}\cdot\left\{\begin{aligned} &\frac{1}{2}\cdot C_1\cdot\tau_{straggle}\cdot\left(\sqrt{\pi}\cdot\left(1+erf\left(\frac{\tilde{R}_{csda}}{\tau_{straggle}}\right)\right)+\tau_{straggle}^{\;2}\cdot e^{-\frac{\tilde{R}_{csda}^{\;2}}{\tau_{straggle}^{\;2}}}\right) + C_2\cdot\tilde{R}_{csda}^{\;2} \\ &+2\cdot C_3\cdot\left(\frac{z_{max}}{P_E\cdot\pi}\right)^2\cdot e^{-\frac{PE\cdot\pi\cdot R_{csda}}{z_{max}}}\left[\frac{P_E\cdot\pi\cdot\tilde{R}_{csda}}{z_{max}}\cdot e^{\frac{P_E\cdot\pi\cdot\tilde{R}_{csda}}{z_{max}}} + 1 - e^{\frac{P_E\cdot\pi\cdot\tilde{R}_{csda}}{z_{max}}}\right] \\ &+C_4\cdot\frac{\tilde{R}_{csda}^{\;2}}{R_{csda}^{\;2}}\left(\frac{1}{2}\cdot\tilde{R}_{csda}^{\;2} + z_{shift}^{\;2} - \frac{4}{3}\cdot\tilde{R}_{csda}\cdot z_{shift}\right) \\ &+C_{Lan1}\cdot E_0^{-1}\cdot\tilde{R}_{Lan1}^{\;2} - C_{Lan2}\cdot\frac{\tilde{R}_{Lan2}^{\;3}}{R_{Lan2}^{\;2}}\cdot\left(\frac{1}{2}\cdot R_{Lan2} + \frac{17}{6}\cdot z_{shift}\right) \end{aligned}\right\} \quad (L45)$$

$$\left.\begin{aligned}\tilde{R}_{csda} &= R_{csda} - z_{shift}\\ \tilde{R}_{Lan1} &= R_{Lan1} - z_{shift}\\ \tilde{R}_{Lan2} &= R_{Lan2} - z_{shift}\end{aligned}\right\} \quad \text{(L46)}$$

$A_{rp}$ and $A_{sp,r}$ yield a rather complicated equations; they can be simplified by application of the mean value theorem of the integral calculus:

$$\left.\begin{aligned}A_{rp} &= C_{heavy}\cdot(R_{csda} - \sqrt{2}\cdot\tau_{heavy} - 3\pi\tau_{Range})\\ A_{sp,r} &= 2\cdot C_{heavy}\cdot\upsilon\cdot\sqrt{\pi}\cdot(R_{csda} - z_{shift}\cdot\sqrt{\pi}) + \tfrac{1}{2}\cdot\sqrt{\pi}\cdot c_1\cdot z_{shift}/R_{csda}\end{aligned}\right\} \quad \text{(L47)}$$

[+]The integration procedure underlying Eq. (L42) is shown by the term $C_2$ of Eq. (14):

$$\left.\begin{aligned}&(\lim z_0\to\infty)\ \ C_2\cdot\int_{-z_0}^{z_0}[1 + erf((R_{csda} - z)/\tau)]\,dz =\\ &= C_2\cdot[2\cdot z_0 - \tau\cdot(u_2\cdot erf(u_2) - u_1\cdot erf(u_1)) + \tfrac{1}{\sqrt{\pi}}\cdot(\exp(-u_2{}^2) - \exp(-u_1{}^2))] \ = 2\cdot C_2\cdot R_{csda}\\ &\qquad u_2 = (R_{csda} - z_0)/\tau;\ u_1 = (R_{csda} + z_0)/\tau;\quad z_0\to\infty\end{aligned}\right\} \quad \text{(L48)}$$

In order to evaluate the integral over $erf((R_{csda} - z)/\tau)$ of formula (L48) we have to perform the substitution $u = (R_{csda} - z)/\tau$ and to use the chain rule:

$$\int erf(u)du = u\cdot erf(u) + \tfrac{1}{\sqrt{\pi}}\cdot\exp(-u^2) \quad \text{(L49)}$$

Since $erf((R_{csda} - z_0)/\tau)$ rapidly assumes -1 for $z_0 \to \infty$ and $erf((R_{csda} + z_0)/\tau) \to 1$, we finally obtain $2\cdot C_2\cdot R_{csda}$ as a contribution to $A_{pp1}$; the Gaussian terms in Eq. (L48) mutually cancel for $z_0 \to \infty$. The evaluation of all other terms leading to equations (L42 – L47) follows the same principle.